\documentclass[12pt]{iopart}

\usepackage{iopams}  
\usepackage[english]{babel}

\newtheorem{theorem}{Proposition}
\newcommand{\eqref}[1]{(\ref{#1})}

\begin{document}

\title{Note on the thermodynamics and the speed of sound of a scalar field}

\author{Oliver~F.~Piattella $^1$, J\'ulio~C.~Fabris $^1$ and Neven Bili\'c $^{2,3}$}
\address{$^1$ Departamento de F\'isica, Universidade Federal do Esp\'irito Santo, Avenida Fernando Ferrari 514, 29075-910 Vit\'oria, Esp\'irito Santo, Brazil}
\address{$^2$ Rudjer Bo\v{s}kovi\'{c} Institute, P.O. Box 180, 10001 Zagreb, Croatia}
\address{$^3$ Departamento de F\'isica, Universidade Federal de Juiz de Fora,\\ 
36036-330, Juiz de Fora, MG, Brazil}
\ead{oliver.piattella@pq.cnpq.br, fabris@pq.cnpq.br, bilic@irb.hr}

\begin{abstract}
We investigate the correspondence between a perfect fluid and a scalar field and show a possible way of expressing thermodynamic quantities such as entropy, particle number density, temperature and chemical potential in terms of the scalar field $\phi$ and its kinetic term $X$. We prove a theorem  which relates isentropy with purely kinetic Lagrangian. As an application, we study the evolution of the gravitational potential in cosmological perturbation theory.
\end{abstract}


\submitto{\CQG}

\maketitle

\section{Introduction}

The discovery of a Higgs-like particle \cite{Aad:2012tfa} strengthens the idea that scalar fields may play a fundamental role in particle physics. Scalar fields are also important in cosmology, since they are at the basis of  most inflationary theories \cite{Linde:1983gd, Lyth:1998xn, ArmendarizPicon:1999rj, Garriga:1999vw, Khlopov:2004rw, DiezTejedor:2005zv, Mazumdar:2010sa, Allahverdi:2010xz, Biswas:2013lna} and are candidates for dark energy \cite{Fabris:2000ee, Copeland:2006wr}. Moreover, non-canonical scalar fields can have in some instances vanishing effective speed of sound, thus being able to model cold dark matter \cite{Matos:1998vk, Fabris:2011wz, Diez-Tejedor:2013sza} or even the entire dark sector of the universe \cite{Padmanabhan:2002sh, DiezTejedor:2006qh, Bertacca:2007ux, Bertacca:2007fc, Bertacca:2008uf, Gao:2009me, Lim:2010yk, Bertacca:2010ct, Bertacca:2010mt, Fabris:2011rm, Daouda:2012ig}. See also \cite{Khlopov:1985jw} for an investigation of the stability of scalar fields and the 
formation of 
primordial black holes.
\par
In the framework of fluid mechanics, it is straightforward to interpret physical quantities such as density, pressure and four-velocity in terms of scalar field quantities \cite{Schutz:1970my, Erickson:2001bq, Babichev:2007dw, Bilic:2008zk, Bilic:2008yr, Bilic:2008zz, Faraoni:2012hn, Semiz:2012zz}. The same can be done, less straightforwardly, for the speed of sound. The
identification usually stops here, since the aforementioned quantities are sufficient to describe, e.g., the
dynamics of a cosmological model. In this paper, we aim to provide a full correspondence between the scalar field and thermodynamic quantities such as entropy per particle, particle number, enthalpy, temperature and chemical potential. Our results are summarized in table~\ref{tab1}. 
\par
We also address the speed of sound, investigating different possible definitions and their counterparts in the scalar field representation. Whereas in fluid mechanics it seems possible to define different physical speeds of sound, in scalar field theory the propagation velocity of small perturbations corresponds just to the adiabatic speed of sound.
\par
The main result of our investigation is a theorem which states that the entropy per particle can be identified with the scalar field itself, except for the case of a purely kinetic Lagrangian. This is particularly relevant in cosmology, where adiabatic models are usually considered. Employing a generic scalar field description thus enforces one to introduce entropic perturbations, except for some special cases, like k-essence models or a scalar field
representing a vacuum state. 

As we will demonstrate,  the equation for the evolution of the gravitational potential in cosmological perturbation theory derived for a general scalar field Lagrangian precisely agrees with the equation derived for a non-adiabatic fluid \cite{Mukhanov:1990me} (see also \cite{Strokov:2006mb} for a discussion on the equivalence of the hydrodynamical and field approaches in the theory of cosmological scalar perturbations of a single medium and \cite{Unnikrishnan:2010ag, Diez-Tejedor:2013nwa} for a discussion on how scalar fields could mimick an isentropic flow.).
\par
We organize the paper as follows. In Sec.~\ref{Sec:SoS} we discuss the definition of the speed of sound in fluid mechanics and 
illustrate it in the example of an ideal Boltzmann gas. In Sec.~\ref{Sec:Framework} we present the field theoretical
description of a perfect fluid and the derivation of the speed of sound. In Sec.~\ref{Sec:TherQuant} we discuss the relation between the scalar field and various thermodynamic quantities. In Sec.~\ref{Sec:AnotherProof} we derive necessary and sufficient conditions that a scalar field Lagrangian must satisfy in order to describe an isentropic fluid. In Sec.~\ref{Sec:AppCosm} we discuss entropic perturbations in a cosmological context and in Sec.~\ref{Sec:Concl} we present our conclusions.

\section{Speed of sound and thermodynamics of a scalar field}\label{Sec:SoS}

In fluid mechanics the speed of sound is defined as a derivative of the pressure with
respect to the density at fixed entropy per particle \cite{landau1987fluid}, i.e.,
\begin{equation}\label{fluisadcs2}
 c_{\rm s}^2 = \left.\frac{\partial p}{\partial\rho}\right|_{s/n} ,
\end{equation}
where $s = S/V$ is the entropy density, $n = N/V$ is the particle number density. 
Since $s/n$ is held fixed, \eqref{fluisadcs2} defines the so-called \textit{adiabatic speed of sound}.
 Physically, the speed of sound is the velocity at which adiabatic compression and rarefaction waves propagate in a medium.

There exists a subtle difference between the concept of \textit{adiabaticity} and \textit{isentropy}. 
Following, for example \cite{Maartens:1996vi}, a flow is said to be \textit{adiabatic}  if the specific entropy is constant along each fluid particle word-line, i.e., if
\begin{equation}\label{adcond}
 (s/n)_{,\mu}u^\mu = 0 ,
\end{equation}
 whereas, a flow is said to be \textit{isentropic} when $s/n$ is the same constant along each word-line, i.e., if
\begin{equation}\label{iscond}
(s/n)_{,\mu} = 0 ,
\end{equation}
or $\rmd(s/n) = 0$. The latter is clearly a stronger requirement than the former since \eqref{iscond} implies
\eqref{adcond} but not the other way around. For a perfect fluid, a sufficient condition for adiabaticity is the energy-momentum conservation $T^{\mu\nu}{}_{;\mu} = 0$ together 
with the particle number conservation $(nu^\mu)_{;\mu} = 0$ \cite{landau1987fluid,Maartens:1996vi,taub1978relativistic,Schutz:1985jx}.

Equation~\eqref{fluisadcs2} suggests that we could, in principle, define various ``speeds of sound'', depending on which quantity we keep fixed while calculating the derivative of $p$ with respect to $\rho$. Let us go into some detail. For a perfect fluid we have in general $\rho, p, s, n$ as grand-canonical thermodynamic functions each depending on two variables: temperature $T$  and chemical potential $\mu$. However, we can trade the functional dependence $p = p(T,\mu)$ for $p = p(\rho,Y)$, where  $Y$ is $s$, $n$, $T$, or any of their combinations. Then, a small departure of pressure from its equilibrium value can be written as
\begin{equation}
 \delta p(\rho, Y) = \left.\frac{\partial p}{\partial \rho}\right|_Y\delta\rho + \left.\frac{\partial p}{\partial Y}\right|_\rho\delta Y .
\end{equation}
In this way we can formally define several ``speeds of sound'' depending on which variable $Y$ is held fixed.
\par
Strictly speaking, \eqref{fluisadcs2} is derived assuming an isentropic irrotational flow or, equivalently, a velocity potential $\psi$ such that $h u_\mu=\psi_{,\mu}$ (for a simple derivation see \cite{Bilic:1999sq}; for details see also \cite{landau1987fluid,taub1978relativistic}). It is far from obvious that a similar derivation could be done for $Y \neq s/n$. In principle, if instead of $\delta(s/n) = 0$ one assumes $\delta Y =0$ and shows that $c_Y^2\equiv\partial p /\partial\rho|_Y$ is related to the velocity coefficient in the perturbation wave equation, then the quantity $c_Y$ would have
the physical meaning of propagation velocity of perturbations at fixed $Y$. 

Before we proceed to study the scalar field description of fluid dynamics, consider as an example a collision-dominated fluid. 

\subsection{Collision-dominated gas}

A collision-dominated gas of particles with rest mass $m$ in equilibrium at nonzero temperature
$T$ can be described  as  
\cite{Maartens:1996vi}
\begin{equation}
 mn = c_0\frac{K_2(\beta)}{\beta} , \qquad p = \frac{mn}{\beta} , 
\qquad \rho = mn\left[\frac{K_1(\beta)}{K_2(\beta)} + \frac{3}{\beta}\right] , 
\end{equation}
where $\beta=m/T$, $c_0$ is a constant, and $K_1$ and $K_2$ are the modified Bessel functions of the second kind. 
For simplicity, we consider the non-relativistic limit $\beta \gg 1$.
Taking into account the asymptotic of the modified Bessel function, the above equations become
\begin{equation}\label{approxbetal}
 p = nT , \qquad \rho \simeq mn + \frac{3}{2}nT .
\end{equation}
Now, if we choose $\rho$ and $n$ as independent thermodynamic variables, then the equation of state for the pressure is
\begin{equation}
 p  \simeq \frac{2}{3}\rho - \frac{2}{3}mn , \qquad \left.\frac{\partial p}{\partial\rho}\right|_n = \frac{2}{3} ,
\end{equation}
so the ``speed of sound" at fixed particle number is $2/3$. 
On the other hand, if we choose $\rho$ and $T$ as independent thermodynamic variables, we find
\begin{equation}
 p  \simeq \frac{\rho}{\beta + 3/2}  \simeq \frac{\rho}{\beta} , \qquad \left.\frac{\partial p}{\partial\rho}\right|_T = \frac{1}{\beta + 3/2}  \simeq \frac{1}{\beta} ,
\end{equation}
so the ``speed of sound" at fixed temperature is constant proportional to the temperature, which is vanishingly small due to the assumed $\beta \gg 1$.

Finally, to compute the adiabatic speed of sound, we use the following form of Gibbs' relation
\begin{equation}\label{Gibbsrel2}
 \rmd\left(\frac{\rho + p}{n}\right) = T\rmd\left(\frac{s}{n}\right) + \frac{1}{n}\rmd p ,
\end{equation}
which, together with \eqref{approxbetal} allows us to write down an explicit relation between the particle number and the temperature when isentropy is assumed, i.e., when $\rmd\left(s/n\right) = 0$:
\begin{equation}\label{ntrel}
 \frac{\rmd n}{n} = \frac{3}{2}\frac{\rmd T}{T} .
\end{equation}
Using this we can recast \eqref{approxbetal} in the form
\begin{equation}
 p = kT^{5/2}, \qquad \rho = kT^{3/2}\left(m + \frac{3}{2}T\right) ,
\end{equation}
where $k$ is the integration constant of \eqref{ntrel}. 
Then one finds
\begin{equation}
 \left.\frac{\partial p}{\partial\rho}\right|_{s/n} = \frac{5}{3}\frac{1}{\beta + 5/2}  \simeq \frac{5}{3\beta}
\end{equation}
as the adiabatic speed of sound, which vanishes in the limit $\beta \rightarrow \infty$.

\section{Field-fluid correspondence and the\\ speed of sound}
\label{Sec:Framework}
 
In this section we briefly review the scalar field theory description of a perfect fluid
and derive the adiabatic speed of sound that corresponds to the standard fluid definition
\eqref{fluisadcs2}. Although the concepts presented here are well-known 
(see, e.g., \cite{Faraoni:2012hn,Tsagas:2007yx}), we give this short review for the sake of completeness and for pedagogical reasons. 

Consider a scalar field $\phi$ whose dynamics 
in a generic spacetime of geometry $g_{\mu\nu}$ (with signature $+,-,-,-$)
is described by the action
\begin{equation}
 S = \int \mathcal{L}\left(X,\phi\right)\sqrt{-g} \rmd^4x ,
\label{eq1}
\end{equation}
where the Lagrangian
$\mathcal{L}\left(X,\phi\right)$ is  an arbitrary
function of $\phi$ and the kinetic term 
\begin{equation}
 X= g^{\mu\nu}\phi_{,\mu}\phi_{,\nu}/2 .
\label{eq01} 
\end{equation}
Variation of the action
with respect to the metric leads to the stress tensor
\begin{equation}\label{scalfielT}
 T_{\mu\nu} = \mathcal{L}_X\phi_{,\mu}\phi_{,\nu} - \mathcal{L}g_{\mu\nu},
\end{equation}
where $\mathcal{L}_X = \partial \mathcal{L}/\partial X$. 
For $X > 0$ the stress tensor \eqref{scalfielT} may be expressed in the form of a perfect fluid: 
\begin{equation}\label{perfectfT}
 T_{\mu\nu} = \left(\rho + p\right)u_{\mu}u_{\nu} - pg_{\mu\nu} ,
\end{equation}
where the fluid functions and velocity are  identified as
\begin{equation}\label{fluidfield}
 p = \mathcal{L},   \quad \rho = 2X\mathcal{L}_X - \mathcal{L}, \quad u_\mu = \frac{\phi_{,\mu}}{\sqrt{2X}}, 
\end{equation}
 Note that the four-velocity is properly normalized as $u_\mu u^\mu = 1$.

For illustrative purposes we will consider in the following  a canonical scalar field 
Lagrangian 
\begin{equation}\label{eq100}
 \mathcal{L}(X,\phi) = X - V(\phi),   
\end{equation}
in the context of Friedmann-Robertson-Walker (FRW) cosmology with line element
\begin{equation}
 \rmd s^2=\rmd t^2 - a(t)^2 (\rmd r^2 + r^2 \rmd\Omega^2).
 \label{eq102}
\end{equation}
In this case we have $\mathcal{L}_{X} = 1$, 
$\rho = X + V$ and
$p = X - V$.
The relevant evolution equations are the field equation  
\begin{equation}
 \ddot{\phi}+3H\dot{\phi}+V_\phi=0,
 \label{eq103}
\end{equation}
the Friedmann equation
\begin{equation}
\rho= X+V=\frac{3}{8\pi G}H^2
 \label{eq104}
\end{equation}
and the continuity equation
\begin{equation}
 \dot{\rho}+3H(p+\rho)=0 ,
 \label{eq105}
\end{equation}
where $H=\dot{a}/a$ is the Hubble expansion rate which is assumed to be positive. Note that \eqref{eq103} can be obtained from \eqref{eq105} via the identification \eqref{fluidfield}.

\subsection{Speed of sound}

In this section we briefly review the derivation of the adiabatic speed of sound \cite{Bilic:2008yr}
and the so called {\em effective speed of sound} \cite{Garriga:1999vw}.
The functions $p = p(X,\phi)$ and $\rho = \rho(X,\phi)$ cannot in general be inverted in order 
to give a closed barotropic form $p=p(\rho)$. However, assuming that $X$ and $\phi$ are independent variables we may  
define the adiabatic speed of sound as the ratio of the total differentials $\rmd p$ and $\rmd\rho$ subject
to the constraint that the entropy per particle remains constant. 
More explicitly, we can write \eqref{fluisadcs2} as
\begin{equation}\label{gensos}
 c_{\rm s}^2 = \left.\frac{p_{X}\rmd X + p_{\phi}\rmd\phi}{\rho_{X}\rmd X + \rho_{\phi}\rmd\phi}\right|_{s/n} ,
\end{equation}
and calculate the differentials imposing the constraint of isentropy. Here and from here on, the subscripts
$X$ and $\phi$ denote  the derivatives with respect to $X$ and $\phi$, respectively. Isentropy implies  constraints 
on the total differentials. 
In particular, as we will shortly demonstrate, $\rmd(s/n)=0$ implies $\rmd\phi=0$.

For an isentropic and irrotational flow, as a consequence of \eqref{fluidfield} we identify the flow potential with the scalar field $\phi$ itself:
\begin{equation}\label{keypoint}
 h u_\mu = \phi_{,\mu} ,
\end{equation}
where
\begin{equation}
h = \frac{\rho + p}{n} 
\label{eq02}
\end{equation}
is the  enthalpy. Comparing \eqref{keypoint} with the definition of the four-velocity in \eqref{fluidfield} 
one reads out 
\begin{equation}\label{enthalpy}
 h = \sqrt{2X} .
\end{equation}
From this and the definition \eqref{eq02} together with \eqref{fluidfield} 
we obtain an expression for the particle density in terms of the scalar field quantities:
\begin{equation}\label{nisentropic}
 n = \sqrt{2X}\mathcal{L}_X .
\end{equation}
Demanding $\rmd(s/n) = 0$  one can rewrite \eqref{Gibbsrel2} with the help of \eqref{enthalpy}, as
\begin{equation}
 \rmd p = \mathcal{L}_X\rmd X .
\end{equation}
In this equation 
$\rmd p$ is the total differential and hence we must impose $\rmd\phi = 0$ for an isentropic process. Therefore, the constraint $\rmd\phi = 0$ is equivalent to $\rmd(s/n) = 0$ and hence  it is natural to associate 
the field variable $\phi$ with the entropy per particle $s/n$.

\par
It is important to stress that the constraint $\rmd\phi = 0$ describing an isentropic process does not
 mean that $\phi$ is a constant. This would imply $X = 0$ and the theory under investigation would be rather trivial. 
In our considerations, the constraint $\rmd\phi = 0$ means that the total differentials $\rmd p$ 
and $\rmd\rho$ should be calculated by keeping the field variable $\phi$ fixed while allowing $X$ to vary arbitrarily,
 i.e., effectively as if the Lagrangian was a function of $X$ only.
 
 Setting $\rmd\phi = 0$ in \eqref{gensos}, the adiabatic speed of sound becomes
\begin{equation}\label{adiabaticgensos}
 c_{\rm s}^2 = \frac{p_{X}}{\rho_{X}} = \frac{\mathcal{L}_{X}}{\mathcal{L}_{X} + 2X\mathcal{L}_{XX}}.
\end{equation}
For the canonical scalar field described by (\ref{eq100}), 
the equation of state generally cannot be written in the barotropic form
$p=p(\rho)$ which indicates deviations from isentropic behaviour.
The isentropic case is realized by a constant potential term, $V=V_0$.
Under this condition one has $p = \rho-2V_0$ as a special case of a barotropic 
equation of state, which gives the adiabatic speed of sound $c_{\rm s}=\partial p/\partial \rho=1$.
However, it follows from (\ref{adiabaticgensos}) 
that  $c_{\rm s}=1$ even if the potential $V(\phi)$ is a nontrivial function
of $\phi$.

Another example particularly important for cosmology is the string-theory inspired tachyon condensate Lagrangian \cite{Sen:2002an, Sen:2002nu, Sen:2002in, Gibbons:2003gb} 
  \begin{equation}
 \mathcal{L} = -U(\phi)\sqrt{1 - X},
 \label{tachyon}
\end{equation}
which has been suggested as a model for dark matter/energy unification \cite{Padmanabhan:2002sh, Bilic:2008yr}. From \eqref{adiabaticgensos} one obtains
\begin{equation}
 c_{\rm s}^2 = 1 - X .
\end{equation}
By making use of $p = \mathcal{L}$ and $\rho = U/\sqrt{1 - X}$, this can be expressed  as
\begin{equation}
 c_{\rm s}^2 = - \frac{p}{\rho} = -w ,
\end{equation}
where $w$ is the equation of state parameter.

A cosmological model based on a canonical scalar field is not successful from the 
point of view of structure formation because its speed of sound is always equal to the speed of light. 
A unified model based on the tachyon Lagrangian has a positive and small speed of sound 
whereas the equation of state parameter is negative, therefore providing a source for an accelerated expansion
(for a review on the subject see \cite{Bertacca:2010ct}).

Note that there is no loss of generality in our choice of the scalar field as the flow potential, in \eqref{keypoint}. 
Indeed, one could in principle use a more general expression for the velocity potential:
\begin{equation}\label{redfield}
 \tilde{h}u_\mu = \tilde\phi_{,\mu},
\end{equation}
where
\begin{equation}\label{transphi}
 \tilde\phi_{,\mu} = f(\phi)\phi_{,\mu},
\end{equation}
with $f(\phi)$ being an arbitrary function of $\phi$. Under this transformation one has
\begin{equation}\label{transX}
 \tilde{X} = \frac{1}{2}g^{\mu\nu}\tilde{\phi}_{,\mu}\tilde{\phi}_{,\nu} = f(\phi)^2X,
\end{equation}
and therefore
\begin{equation}
 \tilde{h} = \sqrt{2\tilde{X}} = f(\phi)\sqrt{2X}.
\end{equation}
Equation~\eqref{transphi} is just a field redefinition under which the Lagrangian transforms as
\begin{equation}\label{Ltilde}
 \mathcal{L} \to \tilde{\mathcal{L}}(\tilde{X},\tilde{\phi}) = \mathcal{L}(f^2X,\phi).
\end{equation}
Even though $\tilde{\mathcal{L}}$ as a function of $(\tilde{X},\tilde{\phi})$ is different from $\mathcal{L}$ as function of $(X,\phi)$, \eqref{Ltilde} implies that $p$ and $\rho$ are invariant under the transformation~\eqref{transphi}. In other words the transformation~\eqref{transphi} serves as a reparametrization of the equation of state. From \eqref{eq02} and \eqref{adiabaticgensos} it is straightforward to show that $nh$ and $c_{\rm s}^2$ are also invariant under \eqref{transphi}.

Physically, the adiabatic speed of sound is the propagation velocity of scalar field perturbations. The result derived  here using thermodynamic and hydrodynamic arguments coincides with the effective speed of sound derived from the wave equation that describes the dynamics of field perturbations \cite{Garriga:1999vw, Kang:2007vs}. 
This may be seen as follows. Consider a general Lagrangian ${\cal L}(X,\phi)$, and write the field as
\begin{equation}
\phi(x)\rightarrow \phi(x) +\chi(x) ,
\label{eq2}
\end{equation}
where $\chi$ is a small perturbation. To derive the wave equation for $\chi$ we expand the Lagrangian up to the quadratic terms in $X$ and keep only the terms quadratic in the derivatives of $\chi$:
\begin{eqnarray}
{\cal L}(X,\phi,\chi) &=& \frac{1}{2}{\cal L}_X g^{\mu\nu}\chi_{,\mu}\chi_{,\nu}
+\frac{1}{2}{\cal L}_{XX}(g^{\mu\nu}\phi_{,\mu}\chi_{,\nu})^2 + \dots \nonumber\\
&=& \frac{1}{2}\left({\cal L}_X g^{\mu\nu} + 2X{\cal L}_{XX} u^\mu u^\nu\right) \chi_{,\mu}\chi_{,\nu} + \dots.
\label{eq3}
\end{eqnarray}
The linear perturbations $\chi$ propagate in the effective metric 
\begin{equation}
G^{\mu\nu}\propto {\cal L}_X g^{\mu\nu}
+  2X{\cal L}_{XX} u^\mu u^\nu ,
 \label{eq4}
\end{equation}
and the propagation is governed by the equation of motion
\begin{equation}
\frac{1}{\sqrt{-g}}
\partial_{\mu}
\left[{\sqrt{-g}}\,( {\cal L}_X \, g^{\mu\nu}+ 2X{\cal L}_{XX}\,u^{\mu}u^{\nu}) \partial_{\nu}\chi\right] + \dots = 0 .
\label{eq5}
\end{equation}
In the comoving reference frame in flat background, (\ref{eq5}) reduces to the  wave equation
\begin{equation}
(\partial_t^2 - c_{\rm s}^2\nabla^2 + \dots)\chi = 0 ,
\label{eq014}
\end{equation}
with the effective speed of sound $c_{\rm s}$ given by (\ref{adiabaticgensos}).

\section{Thermodynamic quantities and scalar field}\label{Sec:TherQuant}

In this section we express other thermodynamic quantities, such as the temperature $T$ 
and the chemical potential $\mu$
in terms of $X$ and $\phi$. 
Besides, we determine the formal expression for the ``speed of sound" when the temperature 
or the particle number density is held fixed. 

We have seen that $\rmd(s/n) = 0$ implies $\rmd\phi = 0$ and therefore we identify 
\begin{equation}
\phi =\frac{s}{n},
\label{eq10}
\end{equation}
together with \eqref{nisentropic}, which has been derived under the assumption $\rmd(s/n)= 0$. Note that we could have assumed a more general relation between 
the field $\phi$ and the entropy per particle $s/n$, e.g. $\phi = F(s/n)$, 
where $F$ is a generic function with $F' \neq 0$, but \eqref{eq10} is the most economical.
Here we assume that \eqref{nisentropic} is a valid
definition for the particle number density related to the action \eqref{eq1} even if $\rmd(s/n)\neq 0$. 
This assumption is supported by the field equation
\begin{equation}
 \left(\mathcal{L}_Xg^{\mu\nu}\phi_{,\nu}\right)_{;\mu} = \mathcal{L}_\phi ,
 \label{field}
 \end{equation}
which may be written as 
\begin{equation}
 \left(n u^{\mu}\right)_{;\mu} = \mathcal{L}_\phi ,
\end{equation}
with $n$ given by \eqref{nisentropic}. 
This equation expresses a nonconservation of the current 
$J^\mu= n u^\mu$ if the right-hand side does not vanish. 
If the entropy per particle is conserved, then the variable $\phi$ in the Lagrangian 
should be kept constant and the right-hand side of \eqref{field} vanishes 
leaving a conservation equation for $J^\mu$. Hence, the particle number $n$ 
defined in \eqref{nisentropic} is conserved for an isentropic process, 
but for a more general one is not. If the entropy per particle is not conserved, 
then it is natural to expect that $n$ will not be conserved either. See also \cite{DiezTejedor:2005fz} for a discussion on hydrodynamics with non-conserved number of particles and how it can be modelled with effective fluid Lagrangian which explicitly depend on the velocity potentials.

Nonconservation of particle number and deviation from isentropy is typical of
relativistic statistical ensemble of neutral bosons at finite temperature. In such a system
there is no distinction
between particles and  antiparticles, and hence, particle creation or annihilation is allowed as long as 
the  conservation of energy is respected. The situation here is very similar.
To illustrate this, consider again the canonical scalar field Lagrangian (\ref{eq100})
which generally respects neither particle number nor entropy per particle conservation.
In this example, the ``particle number" density $n=\sqrt{2X}$ is basically  the square root of
the kinetic energy density. 
The nonconservation of particle number reflects the fact that 
 kinetic energy alone is not a constant of motion. Conservation of energy requires 
  the constancy of the sum $X+V+$ gravitational energy. 
In the cosmological context this means
  the constraint $\dot{X}+\dot{V}+6HX=0$, 
  which follows from the continuity equation (\ref{eq105}). If the potential were independent
of $\phi$, the quantity $n$ would be conserved and equation (\ref{field}) would be equivalent to the 
continuity equation.

Deviation from isentropy does not break the parallel with a perfect fluid because the latter 
admits a situation in which the entropy is not conserved, e.g.,
when bulk viscosity is present \cite{Sussman:1994uy, Zimdahl:1996ka, Piattella:2011bs}. See also \cite{Pujolas:2011he} for a discussion of the identification of the bulk viscosity contribution in non-canonical scalar field theories with kinetic gravity braiding.
\par
In order to complete the identification correspondence between fluid and scalar field quantities, we start from the grand-canonical thermodynamic identity
\begin{equation}
Ts=p+\rho- \mu n ,
 \label{eq03}
 \end{equation}
which may be written in the form of a Legendre transformation from $\rho$ to $p$:
\begin{equation}
 p(T,\mu)= -\rho(s,n) +T s + \mu n ,
 \label{eq04}
 \end{equation}
where the variables are subject to the conditions
\begin{equation}
 s=\frac{\partial p}{\partial T}, \quad n=\frac{\partial p}{\partial \mu},
 \label{eq05}
 \end{equation}
\begin{equation}
 T=\frac{\partial \rho}{\partial s}, \quad \mu=\frac{\partial \rho}{\partial n} .
 \label{eq06}
 \end{equation}
From \eqref{eq06} with the help of \eqref{fluidfield}, \eqref{eq10} and \eqref{nisentropic}
the temperature is calculated as  
\begin{equation}
 T=\left. \frac{\partial \rho}{\partial X} \frac{\partial X}{\partial s}\right|_n
 +\left. \frac{\partial \rho}{\partial \phi} \frac{\partial \phi}{\partial s}\right|_n
 =-\frac{\mathcal{L}_\phi}{\sqrt{2X}\mathcal{L}_X}.
 \label{eq07}
 \end{equation}
Similarly, the chemical potential may be obtained either from \eqref{eq06} or directly from \eqref{eq04}.
Either way we find
\begin{equation}\label{musc}
 \mu = \frac{2X\mathcal{L}_X + \mathcal{L}_\phi \phi}{\sqrt{2X}\mathcal{L}_X}.
\end{equation}
Another way to obtain these results is to start from Gibbs' relation \eqref{Gibbsrel2} and substitute in it the known expressions for $\rho$ and $n$. It is straightforward to find
\begin{equation}
 \left(T\sqrt{2X}\mathcal{L}_X + \mathcal{L}_\phi\right)\rmd\phi = 0,
\end{equation}
and therefore it is natural to identify
\begin{equation}
 T = -\frac{\mathcal{L}_\phi}{\sqrt{2X}\mathcal{L}_X}.
 \label{eq101}
\end{equation}
For the chemical potential $\mu$, substituting the the known expressions for $p$, $\rho$ and $n$ and the new result for the temperature in \eqref{eq03}, one recovers \eqref{musc}. Note that the temperature and entropy defined by \eqref{eq07} and \eqref{eq10}, respectively, may take negative values!
This should not be regarded as unphysical since the temperature and entropy discussed here are thermodynamic analogues with no usual physical meaning as in a thermal ensemble. The temperature and entropy associated with thermal fluctuations
could be studied using the finite temperature Euclidean partition function corresponding to the action 
\eqref{eq1} \cite{Bilic:2008zk}.

Obviously, in view of (\ref{field}) and (\ref{eq07})
there is a relation between the temperature and particle creation/annihilation.
To illustrate this consider our canonical example (\ref{eq100}) in the context
of FRW cosmology.
According to (\ref{field}) if 
$\mathcal{L}_{\phi} \neq 0$, the comoving particle number density
$a^3n$  changes with time.
In other words, there is a particle creation (annihilation) whenever 
$\mathcal{L}_{\phi}> 0$ ($<0$).
On the other hand 
the temperature defined in (\ref{eq07}) measures the slope of the potential $V_\phi$,
so $T<0$ ($>0$) 
corresponds to particle creation (annihilation).

Since the ``particle number" density $n= \sqrt{2X}$
is proportional to the square root of the kinetic energy density, 
  kinetic energy should also change
  as a consequence of (\ref{field}) and $\mathcal{L}_{\phi}\neq 0$.
  This change must be consistent with the slope of the potential because
 a positive (negative) slope of the potential corresponds to
  decreasing (increasing) kinetic energy with $\phi$.
To check the consistency, 
 assume for definiteness $\mathcal{L}_\phi <0$ which implies
$\dot{X}+6HX<0$.
Then, by the continuity equation (\ref{eq105}) $\dot{V}>0$  
which in turn implies  $\dot{\phi} >0$ because  $V_\phi >0$ by the assumption.
This also implies that the kinetic energy density decreases with time which is consistent 
with $\dot{X}+6HX<0$.
Similarly, $\mathcal{L}_\phi >0$  would imply $\dot{V}<0$ 
which together with $V_\phi <0$ and  $\dot{\phi}>0$ 
would yield kinetic energy increasing with time, consistent with $\dot{X}+6HX>0$. 

We summarize our results in the following table:
\begin{table}[htbp]
\centering
\begin{tabular}{|c|c|}
\hline
 Fluid variable & Scalar field counterpart\\
 \hline
 $p$ & $\mathcal{L}$\\
 \hline
 $\rho$ & $2X\mathcal{L}_X - \mathcal{L}$\\
 \hline
 $u^\mu$ & $\phi^{,\mu}/\sqrt{2X}$\\
 \hline
 $s/n$ & $\phi$\\
 \hline
 $n$ & $\sqrt{2X}\mathcal{L}_X$\\
 \hline
 $h$ & $\sqrt{2X}$\\
 \hline 
 $T$ & $ -\mathcal{L}_\phi/(\sqrt{2X}\mathcal{L}_X)$\\
 \hline
 $\mu$ & $(2X\mathcal{L}_X + \mathcal{L}_\phi \phi)/(\sqrt{2X}\mathcal{L}_X)$\\
\hline
 \end{tabular}
\caption{Identification scheme between fluid and scalar field quantities.}
\label{tab1}
\end{table}\\
As an application, we now compute the speed of sound for $n$ held fixed. From the condition $\rmd n = 0$ we obtain:
\begin{equation}
 \rmd X\left(\mathcal{L}_X + 2X\mathcal{L}_{XX}\right) = -2X\mathcal{L}_{X\phi}\rmd\phi ,
\end{equation}
and calculating with this constraint
\begin{equation}\label{nsos}
 c_{\rm n}^2 = \left.\frac{p_{X}\rmd X + p_{\phi}\rmd\phi}{\rho_{X}\rmd X + \rho_{\phi}\rmd\phi}\right|_{n} ,
\end{equation}
one finds
\begin{equation}\label{nsos2}
 c_{\rm n}^2 = \frac{2X\mathcal{L}_{X\phi}}{\mathcal{L}_\phi}\frac{\mathcal{L}_X}{\mathcal{L}_X + 2X\mathcal{L}_{XX}} - 1 = \frac{2X\mathcal{L}_{X\phi}}{\mathcal{L}_\phi}c_{\rm s}^2 - 1 .
\end{equation}
For the case of constant enthalpy the calculation is even simpler because $\rmd h = 0$ implies $\rmd X = 0$ and therefore:
\begin{equation}\label{hsos}
 c_{\rm h}^2 = \left.\frac{p_{\phi}\rmd\phi}{\rho_{\phi}\rmd\phi}\right|_{n} = \frac{\mathcal{L}_{\phi}}{2X\mathcal{L}_{X\phi} - \mathcal{L}_\phi},
\end{equation}
Similar calculations could be performed for the cases in which $T$ or $\mu$ are held fixed.

\section{Isentropy and k-essence}\label{Sec:AnotherProof}

We now prove that an isentropic fluid is necessarily described by a purely kinetic k-essence,
i.e, by a Lagrangian that depends only on $X$.

In the following we shall assume the energy-momentum conservation and isentropy; the conservation of the particle number will come out as a byproduct.

\begin{theorem}
 Let the Lagrangian
$\mathcal{L}=\mathcal{L}(X,\phi)$ describe the dynamics of an irrotational ``fluid" flow.
Then the flow will be isentropic if and only if there exist  a field redefinition 
$\tilde{\phi} =\tilde{\phi} (\phi)$ such that 
the transformed Lagrangian is a function of the kinetic term
$\tilde{X}= g^{\mu\nu }\tilde{\phi}_{,\mu}\tilde{\phi}_{,\nu}$
with no explicit  dependence on $\tilde{\phi}$, 
i.e., if and only if the original Lagrangian is equivalent to 
$\mathcal{L}=\mathcal{L}(\tilde{X})$. 
\end{theorem}

{\bf Proof}

We first prove that $\mathcal{L}=\mathcal{L}(X)$ implies $\rmd(s/n)=0$. 
The Lagrangian that depends only on $X$ yields conservation of the current
$J_\mu=\sqrt{2X} \mathcal{L}_X u_\mu$ and hence we may identify $n=\sqrt{2X} \mathcal{L}_X$ as in 
\eqref{nisentropic}. Applying this expression for $n$, and the expressions
\eqref{fluidfield} for $p$ and $\rho$ to the Gibbs identity \eqref{Gibbsrel2}
we find $T\rmd(s/n)=0$. Then, if  $T\neq 0$ we must have $\rmd(s/n)=0$. If $T=0$, by the third law of thermodynamics
$s\equiv 0$, and hence $\rmd(s/n)=0$. 

To  prove the reverse, assume that $\rmd(s/n)=0$ and $\mathcal{L}=\mathcal{L}(X,\phi)$ to be an unknown function
of independent variables $X$ and $\phi$.  
Furthermore, we assume $p$ and $\rho$ to be related to $\mathcal{L}$ as in \eqref{fluidfield},
but the particle number density $n$ to be unknown. 

Consider the Gibbs relation \eqref{Gibbsrel2} with $\rmd(s/n) = 0$. Then, \eqref{Gibbsrel2} may be written as follows:
\begin{equation}\label{Gibbsrel3}
 \rmd\rho - \left(\rho + p\right)\frac{\rmd n}{n}=0 .
\end{equation}
Inserting  $\rmd n = n_{X}\rmd X + n_{\phi}\rmd\phi$ and the total derivative of $\rho$ calculated using 
$\rho= 2X\mathcal{L}_X-\mathcal{L}$ we obtain
\begin{equation}
\fl 
\left(\mathcal{L}_X + 2X\mathcal{L}_{XX} - 2X\mathcal{L}_X\frac{n_{X}}{n}\right) \rmd X
+\left(2X\mathcal{L}_{X\phi} - \mathcal{L}_\phi - 2X\mathcal{L}_X\frac{n_{\phi}}{n}\right) \rmd\phi=0 .
\label{eq30}
\end{equation}
Since  the differentials $\rmd X$ and $\rmd\phi$ are independent by assumption, the above equation will hold
if and only if the coefficients in front of each differential identically vanish. This yields 
two first order differential 
equations for $n$
\begin{equation}
 \frac{n_{X}}{n} = \frac{\left(\sqrt{2X}\mathcal{L}_X\right)_{X}}{\sqrt{2X}\mathcal{L}_X} ,
 \label{eq31}
\end{equation}
\begin{equation}
\frac{n_{\phi}}{n}=\frac{2X\mathcal{L}_{X\phi} - \mathcal{L}_\phi}{2X\mathcal{L}_X} .
\label{eq32} 
\end{equation}
The general solution to \eqref{eq31} may be expressed as
\begin{equation}\label{nsolution}
 n = \rme^{f(\phi)}\sqrt{2X}\mathcal{L}_X ,
\end{equation}
where $f(\phi)$ is an arbitrary function of $\phi$.
 The solution must be consistent with \eqref{eq32}, so upon substitution we obtain
\begin{equation}\label{diffeqLagr}
 -\mathcal{L}_\phi = 2X\mathcal{L}_{X}f_\phi .
\end{equation}
In general, \eqref{diffeqLagr} is a first order partial differential equation for $\mathcal{L}$, whose general solution is an arbitrary function 
$\mathcal{L}=\mathcal{L}(\tilde{X})$
of the principal integral \cite{evans2002partial}
\begin{equation}\label{eq33}
 \tilde{X}= \rme^{-2f(\phi)}X .
\end{equation}
Now, we 
introduce a new scalar field $\tilde\phi$ such that
$\tilde\phi_{,\mu} = \rme^{-f(\phi)}\phi_{,\mu}$
so  
$\tilde{X}=\tilde\phi^{,\mu}\tilde\phi_{,\mu}/2$
is the kinetic term corresponding to $\tilde\phi$. 
Hence, the Lagrangian $\mathcal{L}$ is a function of 
the kinetic term  only 
 which was to be shown.  
This completes the proof. It is interesting to note that in \cite{Akhoury:2008nn} the authors perform the same calculations we did in this proof in order to show that stationary configurations imply a shift symmetry, i.e. symmetry under the transformation $\tilde\phi_{,\mu} = \rme^{-f(\phi)}\phi_{,\mu}$.

It may be easily shown that the particle number \eqref{nsolution}
expressed in terms of $\tilde{X}$  takes the usual form
\begin{equation}
n= \sqrt{2\tilde{X}}\mathcal{L}_{\tilde{X}} 
\end{equation} 
and satisfies the conservation equation $(n u^\mu)_{;\mu}=0$.
The conservation of $n$ may also be shown directly from \eqref{nsolution}  
with \eqref{diffeqLagr}.

\section{Scalar field and cosmological perturbations}\label{Sec:AppCosm}

It is worthwhile to discuss a cosmological scalar field in the present context. 
Let us compare the evolution equations for the gravitational potential in the models 
of a non-adiabatic fluid and of a scalar field with Lagrangian $\mathcal{L}(X,\phi)$. 
For details on cosmological perturbations theory, we refer the reader to 
\cite{Garriga:1999vw, Mukhanov:1990me, mukhanov2005physical}.

Consider the following line element in the conformal Newtonian coordinate system:
\begin{equation}\label{pertmet}
 ds^2 = a(\eta)^2\left[\left(1 + 2\Phi\right)d\eta^2 - \left(1 + 2\Psi\right)\delta_{ij}dx^idx^j\right],
\end{equation}
where $\eta$ is the conformal time and $\Phi$ and $\Psi$ are the gravitational potentials. 
To study fluctuations of the field we replace $\phi \rightarrow \phi(\eta) + \delta\phi$, 
where $\phi(\eta)$ is purely time dependent, whereas $\delta\phi$ may also depend on the spatial coordinates. Then, to linear order, we also have 
\begin{equation}
 X \rightarrow X+ \delta X ,
 \label{eqXdeltaX}
\end{equation}
where the background kinetic term on the right hand side is 
$X =\phi^{'2}/2a^2$ 
and its fluctuation
\begin{equation}
 \delta X = \frac{{\phi'}^2}{a^2}\left( \frac{\delta\phi'}{\phi'}-\Phi\right).
 \label{deltax}
\end{equation}
Here and from here on the prime denotes a derivative  with respect to the conformal time. 
Using the replacements
\begin{eqnarray}
 \mathcal{L} \rightarrow \mathcal{L} + \mathcal{L}_X\delta X + \mathcal{L}_\phi\delta\phi ,\\
 \mathcal{L}_X \rightarrow \mathcal{L}_{X} + \mathcal{L}_{XX}\delta X + \mathcal{L}_{X\phi}\delta\phi ,
\end{eqnarray}
we can write down the full stress-energy tensor as
\begin{eqnarray}\label{setensscalfield}
 T^{\mu}{}_\nu = \mathcal{L}_{X}\phi^{,\mu}{\phi}_{,\nu} - 
\mathcal{L}\delta^{\mu}{}_{\nu} + \mathcal{L}_{XX}\delta X\phi^{,\mu}{\phi}_{,\nu} 
+ \mathcal{L}_{X\phi}\delta\phi\phi^{,\mu}{\phi}_{,\nu} \nonumber\\ 
+ \mathcal{L}_{X}\delta\phi^{,\mu}{\phi}_{,\nu} 
+ \mathcal{L}_{X}{\phi}^{,\mu}\delta\phi_{,\nu} - 
\mathcal{L}_X\delta X\delta^{\mu}{}_{\nu} - \mathcal{L}_\phi\delta\phi\delta^{\mu}{}_{\nu}.
\end{eqnarray}
Since $\phi$ is a function of time only, we have $T^{i}{}_{j} = 0$ for $i \neq j$. 
Therefore, even in the perturbed case no anisotropic stresses are present. 
Moreover, Einstein equations tell us that $\Phi = \Psi$ in \eqref{pertmet}.

One can easily realize that the background contribution to \eqref{setensscalfield} is
\begin{eqnarray}
 {T^0}_0 = \rho = \frac{\mathcal{L}_{X}}{a^2}\phi^{'2} - \mathcal{L} ,\\
 {T^0}_i = 0 ,\\
 {T^i}_j = -p\delta^i{}_j = - \mathcal{L}\delta^i{}_j ,
\end{eqnarray}
whereas the perturbation contributions are
\begin{eqnarray}
 \label{deltarho}
 \delta T^{0}{}_0 = \delta\rho = \left(\mathcal{L}_X + 2X\mathcal{L}_{XX}\right)\delta X + 
\left(2X\mathcal{L}_{X\phi} - \mathcal{L}_\phi\right)\delta\phi  
\\
 \delta T^{0}{}_i =  \mathcal{L}_{X}\frac{1}{a^2}\phi'\delta\phi_{,i} = -(\rho + p)\frac{1}{a^2}\delta u_i,\\
 \label{deltap} \delta T^{i}{}_j = - \delta p \delta^i{}_j = 
-\left(\mathcal{L}_X\delta X + \mathcal{L}_\phi\delta\phi\right)\delta^i{}_j .
\end{eqnarray}
From \eqref{fluidfield} and \eqref{deltax}  it follows
\begin{equation}
 \delta u_\mu = a\frac{\delta\phi_{,\mu}}{\phi'} + a\frac{\phi_{,\mu}}{\phi'}
 \left(\Phi - \frac{\delta\phi'}{\phi'}\right),
 \label{deltau4}
\end{equation}
so that the perturbation of the three-velocity is
\begin{equation}
 \delta u_i = a\frac{\delta\phi_{,i}}{\phi'}.
 \label{deltau3}
\end{equation}
This completes the identification of the physical quantities such as density, 
pressure and three-velocity in terms of the scalar field quantities in the 
context of the cosmological perturbation theory.

Following \cite{mukhanov2005physical}, the fluid model leads to the following equation for 
the gravitational potential:
\begin{equation}\label{gravpotfluid}
 \Phi'' + 3\mathcal{H}\left(1 + c_{\rm s}^2\right)\Phi' - c_{\rm s}^2\nabla^2\Phi + 
\left[2\mathcal{H}' + \left(1 + 3c_{\rm s}^2\right)\mathcal{H}^2\right]\Phi = 
4\pi Ga^2\tau\delta\left(\frac{s}{n}\right) ,
\end{equation}
where $\mathcal{H} = a'/a$ and $\delta(s/n)$ is the entropy per particle perturbation. According to our notation, $c_{\rm s}^2$ is the adiabatic speed of sound and 
$\tau = \partial p/\partial (s/n)|_{\rho}$.

For a generic scalar field, let us write down the relevant Einstein equations. Using \eqref{deltarho} 
and \eqref{deltap} we get
\begin{eqnarray}
 \nabla^2\Phi - 3\mathcal{H}\left(\Phi' + \mathcal{H}\Phi\right) =\nonumber\\ 4\pi G a^2\left(\mathcal{L}_X 
+ 2X\mathcal{L}_{XX}\right)\delta X + 4\pi G a^2\left(2X\mathcal{L}_{X\phi} - \mathcal{L}_\phi\right)\delta\phi,\\
 \Phi'' + 3\mathcal{H}\Phi' + \left(2\mathcal{H}' + \mathcal{H}^2\right)\Phi = 
4\pi G a^2\mathcal{L}_X\delta X + 4\pi G a^2\mathcal{L}_\phi\delta\phi,
\end{eqnarray}
where we have written the right-hand sides for the density and the pressure by simply perturbing \eqref{fluidfield}. 
Since, according to our identification scheme, $\phi = s/n$, combining the two equations and eliminating $\delta X$, 
one finds
\begin{eqnarray}\label{gravpotscal}
 \Phi'' + 3\mathcal{H}\Phi' + \left(2\mathcal{H}' + \mathcal{H}^2\right)\Phi - 
\frac{\mathcal{L}_X}{\mathcal{L}_X + 2X\mathcal{L}_{XX}}\nabla^2\Phi + \nonumber\\ 
 3\mathcal{H}\frac{\mathcal{L}_X}{\mathcal{L}_X + 2X\mathcal{L}_{XX}}
\left(\mathcal{H}\Phi + \Phi'\right) =  
 4\pi G a^2\left(\mathcal{L}_\phi - 
\frac{2X\mathcal{L}_X\mathcal{L}_{X\phi}-\mathcal{L}_X \mathcal{L}_\phi}{\mathcal{L}_X + 2X\mathcal{L}_{XX}}
\right).
\end{eqnarray}
Comparing the two evolution equations \eqref{gravpotfluid} and \eqref{gravpotscal} one can identify:
\begin{equation}
 c_{\rm s}^2 = \frac{\mathcal{L}_X}{\mathcal{L}_X + 2X\mathcal{L}_{XX}} , 
\qquad \tau =\mathcal{L}_\phi - 
\frac{2X\mathcal{L}_X\mathcal{L}_{X\phi}-\mathcal{L}_X \mathcal{L}_\phi}{\mathcal{L}_X + 2X\mathcal{L}_{XX}} ,
 \label{cstau}
\end{equation}
and of course $\delta (s/n) = \delta\phi$. 

Note that the speed of sound found here agrees with \eqref{adiabaticgensos} and 
 that the expression for  $\tau$ in \eqref{cstau} may also be obtained directly from the definition
$ \tau = \partial {\mathcal{L}}/\partial\phi|_\rho$. 

It is important to examine the case $\tau = 0$ which is equivalent 
to having adiabatic perturbations.  Imposing $\tau = 0$, from \eqref{cstau} 
we obtain 
\begin{equation}
 \mathcal{L}_\phi\mathcal{L}_X - X\mathcal{L}_X\mathcal{L}_{X\phi} + X\mathcal{L}_{XX}\mathcal{L}_\phi = 0,
\end{equation}
which can be cast in the form
\begin{equation}\label{Unnikrishnan}
 \left(\frac{\mathcal{L}_\phi}{X\mathcal{L}_X}\right)_X = 0.
\end{equation}
This can be integrated yielding \eqref{diffeqLagr} and confirming the results of the
previous section. The above condition \eqref{Unnikrishnan} was also found in \cite{Unnikrishnan:2010ag} 
in the context of cosmology. However, the authors did not recognize that a Lagrangian
that satisfies the condition \eqref{Unnikrishnan} can always be
put in a purely kinetic form by making use of an appropriate  field transformation.
\par
Consider three particular cases: the kinetic k-essence, the canonical scalar field
and the tachyon condensate. 
In the first case we have $\tau=0$ and obviously the right-hand side of \eqref{gravpotscal}
vanishes as it should since the kinetic k-essence describes an isentropic fluid. 
In  the case of a canonical scalar field (\ref{eq100}), equation
 \eqref{gravpotscal} becomes
\begin{equation}\label{gravpotcanscal}
 \Phi'' + 6\mathcal{H}\Phi' + \left(2\mathcal{H}' + 4\mathcal{H}^2\right)\Phi - \nabla^2\Phi   
   =- 8\pi G a^2V_\phi\delta\phi ,
\end{equation}
which can be found for example in \cite{Mukhanov:1990me}.
Similarly, for the tachyon condensate \eqref{tachyon} we obtain
\begin{eqnarray}\label{gravpottachyon}
 \Phi'' + 3(2-X)\mathcal{H}\Phi' + \left[2\mathcal{H}' + (4-3X)\mathcal{H}^2\right]\Phi - \nabla^2\Phi  = \nonumber
 \\
  - 8\pi G a^2 (1-X)^{1/2}U_\phi\delta\phi .
\end{eqnarray}
Both expressions \eqref{gravpotcanscal} and \eqref{gravpottachyon}
can be cast in a closed form for the gravitational potential by using the $0-i$ Einstein equation
in order to trade $\delta\phi$ for $\Phi$.

It is sometimes convenient to study cosmological perturbations in terms of the density contrast $\delta = \delta\rho/\rho$. 
Following \cite{mukhanov2005physical}, the equations $\delta T^\mu{}_{\nu;\mu} = 0$ may be written as 
\begin{eqnarray}
\label{pertconteq} \delta' = -(1 + w)\left(\theta - 3\Phi'\right) 
- 3\mathcal{H}\delta\left(\frac{\delta p}{\delta\rho} - w\right),\\
 \theta' = -\mathcal{H}\theta\left(1 - 3w\right) - 
\frac{w'}{1 + w}\theta - \frac{\nabla^2\delta p}{\rho + p} - \nabla^2\Phi,
\end{eqnarray}
where $w = p/\rho$ and $\theta = a\delta u^i_{,i}$. Equation  \eqref{pertconteq}  is often expressed in the form
\begin{equation}
 \delta' + 3\mathcal{H}\delta\left(c_{\rm s}^2 - w\right) = -(1 + w)\left(\theta - 3\Phi'\right)
 - 3\mathcal{H}w\Gamma,
\end{equation}
i.e., introducing the adiabatic speed of sound and the so-called entropy perturbation
\begin{equation}
 \Gamma = \left(\frac{\delta p}{\delta\rho} - c_{\rm s}^2\right)\frac{\delta}{w},
\end{equation}
which measures how much the quantity $\delta p/\delta\rho$ departs from the adiabatic speed of sound. 
Using \eqref{fluidfield} and  \eqref{adiabaticgensos} 
 for a generic scalar field we obtain
\begin{equation}
 \Gamma = \frac{\tau}{\mathcal{L}}\delta\phi.
\end{equation}
As expected, the entropy perturbation $\Gamma$ is proportional to $\delta\phi$. 
For a canonical scalar field and the tachyon condensate one finds
\begin{equation}
 \Gamma = \frac{-2V_{\phi}}{\mathcal{L}}\delta\phi,     \quad
 \Gamma = \frac{2U_{\phi}}{U}\delta\phi,
\end{equation}
respectively.

A nonadiabatic scenario has been suggested as a possible way out 
of the structure formation problem immanent to all DM/DE unification  models.
It has been noted by Reis et al. \cite{Reis:2003mw} 
that the root of the structure formation problem is the
term $\nabla^2 \delta p$ in perturbation equations, equal to
$c_{\rm s}^{2} \nabla^2 \delta $ for adiabatic
perturbations, and if there are entropy perturbations such that
$\delta p = 0$, no difficulty arises.
Obviously, this scenario  cannot work for simple 
or generalized Chaplygin gas models
as these models are adiabatic.
As demonstrated in \cite{Bilic:2005sp}, the nonadiabatic scenario also does not work
 for a hybrid Chaplygin gas, which is a two component
nonadiabatic model.

As an application of the formalism we developed here, it would be worthwhile to explore the nonadiabatic scenario in a 
tachyon condensate model \eqref{tachyon} where 
\begin{equation}
 \delta p =p\left( \delta -\Gamma\right)=p\left( \delta -\frac{2U_\phi}{U}\delta\phi\right).
\end{equation}
The nonadiabatic scenario is realized if the entropy perturbations needed to make $\delta p$ vanish
satisfy $2U_\phi\delta\phi=U\delta$
as an initial condition outside the causal horizon
$d_c = \int d\eta \simeq H_0^{-1}  a^{1/2}$ and 
evolve with time in the same way as the density contrast. The evolution equation for $\delta\phi$  which may be derived from \eqref{field}, should be thoroughly analysed
to establish whether a nonadiabatic scenario is possible.
This analysis goes beyond the scope of the present paper and we plan to do it elsewhere.

\section{Conclusions}\label{Sec:Concl}

We have investigated how thermodynamic quantities related to a perfect fluid 
may be expressed in a scalar field theory. The identification of density, 
pressure and four-velocity is straightforward, and widely accepted in the literature, 
once comparing the perfect fluid stress-energy tensor and the stress-energy tensor 
of a scalar field described by a Lagrangian $\mathcal{L}(X,\phi)$. In contrast,
to express the adiabatic speed of sound in terms of the scalar field quantities, one has to 
precisely  define  the notions of isentropy and adiabaticity in scalar field theory. 
We have addressed this point in some detail and proved a theorem which basically states that 
$\rmd(s/n)=0 \Leftrightarrow \rmd(\phi)=0$, leading to the natural identification $\phi = s/n$, for  
a general  scalar field Lagrangian.

We take advantage of Gibbs-Helmholtz and Gibbs-Duhem relations to express 
thermodynamic quantities such as particle number, temperature, enthalpy 
and chemical potential in a scalar field theory, providing a full identification scheme. 
As an application, we have calculated the scalar field equivalent of the speed of sound 
at fixed particle number density.

We have considered the cosmological case and derived the evolution equation for the gravitational potential 
(in the Newtonian gauge) in the presence of a generic scalar field.
Also in this case we have demonstrated a full correspondence between the  field theoretical and the
 fluid hydrodynamic pictures. The adiabatic speed of sound enters the equation with the physical 
meaning of a propagation velocity of adiabatic perturbations. The entropic perturbations correspond to the
 perturbations of the scalar field which supports our identification of the scalar field as the entropy per particle.
 
\ack

OFP and JCF thank CNPq (Brazil) and FAPES (Brazil) for partial financial support. NB thanks CNPq and FAPES for financial support during his visit to the Federal University of Juiz de Fora (UFJF) and to the Federal University of Esp\'irito Santo (UFES) where a part of this work has been completed.

\section*{References}

\bibliographystyle{iopart-num.bst}
\bibliography{SFsos.bib}

\end{document}